\documentclass{webofc}
\usepackage[varg]{txfonts}   % Web of Conferences font
\usepackage{float}
\usepackage{tabularx}
\usepackage{diagbox}
\usepackage{booktabs}
\usepackage{amsmath}
\usepackage{amsfonts}
\usepackage[rgb]{xcolor}
\usepackage{hyperref}
\hypersetup{
	colorlinks=false,
	linkcolor=blue,     
	urlcolor=blue}
\usepackage{epsfig}
\usepackage{lineno}
\usepackage{epstopdf}
\newcommand{\be}{\begin{equation}}
\newcommand{\ee}{\end{equation}}
\newcommand{\ba}{\begin{eqnarray}}
\newcommand{\ea}{\end{eqnarray}}

\newcommand{\la}{\label}

\begin{document}
%%%%%%%%%%%%%%%%%%%%%%%%%%%%%%%%%%%%%%%%%%%%%%%%%%%%%%%%%%%%%%%%%%%%%%
\title{The effective complex heavy-quark potential in an anisotropic quark-gluon plasma}
%%%%%%%%%%%%%%%%%%%%%%%%%%%%%%%%%%%%%%%%%%%%%%%%%%%%%%%%%%%%%%%%%%%%%%%%%%
\author{\firstname{Ajaharul} \lastname{Islam}\inst{1}\fnsep\thanks{\email{aislam2@kent.edu}}
\and
\firstname{Lihua} \lastname{Dong}\inst{2,3,4}\fnsep
%\thanks{\email{donglh6@mail2.sysu.edu.cn}}
\and
 \firstname{Yun} \lastname{Guo}\inst{2,3}\
 %\thanks{\email{yunguo@mailbox.gxnu.edu.cn}}
\and
 \firstname{Alexander} \lastname{Rothkopf}\inst{5}\fnsep
 %\thanks{\email{alexander.rothkopf@uis.no}}
 \and
 \firstname{Michael} \lastname{Strickland}\inst{1}\fnsep
 %\thanks{\email{mstrick6@kent.edu}}
}

\institute{Department of Physics, Kent State University, Kent, OH 44242, United States
\and
           Department of Physics, Guangxi Normal University, Guilin, 541004, China
\and
           Guangxi Key Laboratory of Nuclear Physics and Technology, Guilin, 541004, China
\and
           School of Physics and Astronomy, Sun Yat-Sen University, Zhuhai, 519082, China
\and
           Faculty of Science and Technology, University of Stavanger, Stavanger, NO-4036, Norway
          }
%%%%%%%%%%%%%%%%%%%%%%%%%%%%%%%%%%%%%%%%%%%%%%%%%%%%%%%%%%%%%%%%%%%%%%%%%%%%%%%
\abstract{We introduce a method for reducing anisotropic heavy-quark potentials to isotropic potentials by using an effective screening mass that depends on the quantum numbers $l$ and $m$ of a given state. We demonstrate that, using the resulting 1D effective potential model, one can solve a 1D Schrödinger equation and reproduce the full 3D results for the energies and binding energies of low-lying heavy-quarkonium bound states to relatively high accuracy. This includes the splitting of different p-wave polarizations. The resulting 1D effective model provides a way to include momentum anisotropy effects in open quantum system simulations of heavy-quarkonium dynamics in the quark-gluon plasma.
}
%%%%%%%%%%%%%%%%%%%%%%%%%%%%%%%%%%%%%%%%%%%%%%%%%%%%%%%%%%%%%%%%%%%%%%%%%
\maketitle
%%%%%%%%%%%%%%%%%%%%%%%%%%%%%%%%%%%%%%%%%%%%%%%%%%%%%%%%%%%%
\section{Introduction}
%%%%%%%%%%%%%%%%%%%%%%%%%%%%%%%%%%%%%%%%%%%%%%%%%%%%%%%%%%%%%%
The survival probability of the heavy quarkonium states, such as $J/\Psi$ and $\Upsilon$ has been widely used as a sensitive probe to study the quark-gluon plasma (QGP) formed in relativistic heavy-ion experiments at RHIC and LHC~\cite{Matsui:1986dk,Karsch:1987pv}. Due to non-relativistic nature of heavy quarkonium states, one can obtain their in-medium properties, such as masses and decay rates by solving a Schrödinger equation with a complex heavy-quark (HQ) potential. The real part of the HQ potential provides the binding energy, whereas the imaginary part provides information about the decay of a quarkonium state via wave function decoherence~\cite{Laine:2006ns,Brambilla:2008cx,Beraudo:2007ky,Escobedo:2008sy,Brambilla:2010vq,Brambilla:2011sg,Brambilla:2013dpa}. One can obtain the HQ potential at short distances by making use of hard-thermal-loop (HTL) resummed perturbation theory in the weak-coupling limit. Recently, several attempts have been made to develop complex-valued potential models to understand the in-medium properties of quarkonia quantitatively~\cite{Dumitru:2009ni,Margotta:2011ta,Strickland:2011mw,Strickland:2011aa,Krouppa:2015yoa,Krouppa:2016jcl,Krouppa:2017jlg,Islam:2020gdv,Islam:2020bnp,Wen:2022yjx}. During the last decade, many prior works have treated the QGP as an anisotropic medium by incorporating momentum-space anisotropies generated by longitudinal expansion into the underlying parton distribution functions. To make a phenomenological study of this effect in heavy-ion collisions, we consider the following spheroidal distribution function ansatz in the local rest frame (LRF) of the QGP~\cite{Romatschke:2003ms}
%%%%%%
\be\label{anisodis}
f_{\rm aniso}^{\rm LRF} ({\bf k})\equiv f_{\rm iso}\!\left(\frac{1}{\lambda}\sqrt{{\bf k}^2+\xi ({\bf k}\cdot {\bf n})^2}\right)\,.
\ee
%%%%%%

This form takes into account the rapid longitudinal expansion of the QGP  at early times and allows for explicit pressure anisotropies in the LRF~\cite{Strickland:2014pga,Berges:2020fwq}. Here, $f_{\rm iso}$ is an arbitrary isotropic distribution function, $\lambda$ is a temperature-like scale, which becomes the temperature $T$ of the system in the thermal equilibrium limit. The degree of momentum-space anisotropy ($\xi$) in the range $-1 < \xi < \infty$ is given by 
\begin{equation}
\xi = \frac{1}{2}\frac{\langle \bf k^{2}_{\perp}\rangle}{\langle k^2_z\rangle}-1\, ,
\end{equation}
where $k_z \equiv \bf k \cdot n$ and $\bf k_{\perp}\equiv \bf k-\bf n \, (k\cdot n)$ correspond to the particle momenta along and perpendicular to the direction of anisotropy ($\bf{n}$), respectively. Many prior works have studied heavy quarkonium physics by considering the momentum-space anisotropy inside the QGP~\cite{Strickland:2011aa,Thakur:2013nia,Krouppa:2017jlg,Islam:2020gdv,Islam:2020bnp,Dumitru:2009ni,Romatschke:2003ms,Strickland:2014pga}.  Here we focus on how to efficiently take momentum-space anisotropy into account in a one-dimensional effective theory and compare the one- and three-dimensional results for static and dynamical quantities numerically.

In this proceedings contribution, we summarize our previous works where the real part of a 3D anisotropic HQ potential has been reduced to 1D effective potential~\cite{Dong:2021gnb, Dong:2022mbo}. This work is organized as follow: In sec.~\ref{sec-2} we describe the isotropic complex HQ potential model, in sec. \ref{sec-3} we obtain the anisotropic complex HQ potential model, in sec. \ref{sec-4} we obtain our effective complex HQ potential model, in sec. \ref{sec-5} we present our static results, and in sec. \ref{sec-6} we present our dynamic results.

%%%%%%%%%%%%%%%%%%%%%%%%%%%%%%%%%%%%%%%%%%%%%%%%%%%%%%%%%%%%
\section{Isotropic Potential Model}\label{sec-2}
%%%%%%%%%%%%%%%%%%%%%%%%%%%%%%%%%%%%%%%%%%%%%%%%%%%%%%%%%%%%%%

The Fourier transform of the real time gluon propagator in the static limit gives the complex HQ potential in an isotropic QGP~\cite{Guo:2018vwy}.
%%%
\be\la{revdef}
V (\lambda,r)=-g^2 C_F \int \frac{d^3 {\bf p}}{(2\pi)^3} (e^{i {\bf p} \cdot {\bf r}}-1)D^{00}(p_0=0,{\bf p}, \lambda)\, .
\ee
%%%
%%%%%%%%%%%%%%%%%%%%%%%%%%%%%%%%%%%%%%%%
\subsection{Perturbative Contribution}
%%%%%%%%%%%%%%%%%%%%%%%%%%%%%%%%%%%%%%%%

The perturbative contribution to the complex HQ potential can be obtained from HTL resummed perturbation theory. The real and imaginary parts of this perturbative contribution are given by
%%%
\be\la{revdefpt}
{\rm Re}\,V_{\rm pt} (\lambda,r)=-g^2 C_F \int \frac{d^3 {\bf p}}{(2\pi)^3} (e^{i {\bf p} \cdot {\bf r}}-1)\left(\frac{1}{p^2+m_D^2}-\frac{1}{p^2}\right)\equiv \alpha m_D ({\cal I}_1({\hat r})-1)\, ,
\ee
%%%
\be\la{imvdefpt}
{\rm Im}\,V_{\rm pt} (\lambda,r)=-g^2 C_F \int \frac{d^3 {\bf p}}{(2\pi)^3} (e^{i {\bf p} \cdot {\bf r}}-1)  \frac{-\pi \lambda m_D^2}{p(p^2+m_D^2)^2}\equiv \alpha \lambda ({\cal I}_2({\hat r})-1)\, ,
\ee
%%%
where the integrals ${\cal I}_1({\hat r})$ and ${\cal I}_2({\hat r})$ are
%%%
\ba\la{int1}
{\cal I}_1({\hat r})&=&4\pi \int \frac{d^3 {\hat {\bf p}}}{(2\pi)^3} e^{i {\hat {\bf p}} \cdot {\hat {\bf r}}}\frac{1}{{\hat p}^2({\hat p}^2+1)}=\frac{1-e^{-{\hat r}}}{{\hat r}}\, ,\nonumber \\
{\cal I}_2({\hat r})&=& 4\pi^2\int \frac{d^3 {\hat {\bf p}}}{(2\pi)^3} e^{i {\hat {\bf p}} \cdot {\hat {\bf r}}}\frac{1}{{\hat p}({\hat p}^2+1)^2}= \phi_2({\hat r})\, ,
\ea
%%%
with
\begin{equation}
\phi_n({\hat r}) = 2 \int_0^\infty dz \frac{\mathrm{sin}(z {\hat r})}{z {\hat r}}\frac{z}{(z^2 +1)^n}\, .
\end{equation}
%%%
Here, ${\hat {\bf p}}\equiv{\bf p}/m_D$, $ {\hat {\bf r}}\equiv{\bf r}m_D$, and the strong coupling constant $\alpha=g^2 C_F/(4\pi)$. We also subtracted a term $1/p^2$ in eq.~(\ref{revdefpt}) to make the $r$-independent part finite.

%%%%%%%%%%%%%%%%%%%%%%%%%%%%%%%%%%%%%%%%%%%
\subsection{Non-perturbative Contribution}
%%%%%%%%%%%%%%%%%%%%%%%%%%%%%%%%%%%%%%%%%%%

The gluon propagator also contains a non-perturbative string contribution which arises from a dimension two gluon condensate. Its Fourier transform gives us the non-perturbative contributions~\cite{Guo:2018vwy}
%%%
\be\la{revdefnpt}
{\rm Re}\,V_{\rm npt} (\lambda,r)=-g^2  C_F m_G^2\int \frac{d^3 {\bf p}}{(2\pi)^3} (e^{i {\bf p} \cdot {\bf r}}-1)\frac{p^2+5m_D^2}{(p^2+m_D^2)^3}\equiv -  \frac{2\sigma}{m_D} ({\cal I}_3({\hat r})-1)\, ,
\ee
%%%
\be\la{imvdefnpt}
{\rm Im}\,V_{\rm npt} (\lambda,r)=-g^2 C_F m_G^2 \int \frac{d^3 {\bf p}}{(2\pi)^3} (e^{i {\bf p} \cdot {\bf r}}-1)  \frac{4 \pi \lambda m_D^2(p^2-2m_D^2)}{p(p^2+m_D^2)^4}\equiv  \frac{4\sigma \lambda}{m_D^2}  ({\cal I}_4({\hat r})-1)\, ,
\ee
%%%
where $\sigma=\alpha m_G^2/2$ and $m_G^2$ is a dimensionful constant.  The integrals appearing above are
%%%
\ba\la{int2}
{\cal I}_3({\hat r})&=&4\pi \int \frac{d^3 {\hat {\bf p}}}{(2\pi)^3} e^{i {\hat {\bf p}} \cdot {\hat {\bf r}}}\frac{{\hat p}^2+5}{({\hat p}^2+1)^3}= (1+{\hat r}/2)e^{-{\hat r}}\, ,\nonumber \\
{\cal I}_4({\hat r})&=&8\pi^2\int \frac{d^3 {\hat {\bf p}}}{(2\pi)^3} e^{i {\hat {\bf p}} \cdot {\hat {\bf r}}}\frac{2-{\hat p}^2}{{\hat p}({\hat p}^2+1)^4}=-2\phi_3({\hat r})+6\phi_4({\hat r})\, .
\ea
%%%

%%%%%%%%%%%%%%%%%%%%%%%%%%%%%%%%%%%%%%%%%%%
\subsection{Total Isotropic potential}
%%%%%%%%%%%%%%%%%%%%%%%%%%%%%%%%%%%%%%%%%%%
The sum of the perturbative and non-perturbative contributions give us the total complex isotropic HQ potential
%%%
\ba
\mathrm{Re}\,V_{\mathrm{Iso}} (r) &=& {\rm Re}\,V_{\rm pt} (\lambda,r) + {\rm Re}\,V_{\rm npt} (\lambda,r) \nonumber\\
&=& \alpha m_D \left(\frac{1-e^{-r m_D}}{r m_D}\right)- \alpha m_D - \frac{\sigma}{m_D}\left(2 + r m_D\right)e^{-r m_D} + \frac{2\sigma}{m_D}-\frac{\alpha}{r} \, ,
\ea
%%%
\ba
\mathrm{Im}\,V_{\mathrm{Iso}} (r) &=& {\rm Im}\,V_{\rm pt} (\lambda,r) + {\rm Im}\,V_{\rm npt}  (\lambda,r)\nonumber\\
&=& \alpha \lambda \phi_2\left(r m_D\right) - \alpha \lambda -  \frac{8\sigma\lambda}{m_D^2}\phi_3(r m_D) + \frac{24\sigma\lambda}{m_D^2}\phi_4(r m_D) - \frac{4\sigma\lambda}{m_D^2} \, .
\ea
%%%
We include a relativistic correction, $-0.8 \sigma /(m_{b/c}^2 r)$, in the potential model while solving the Schr{\"o}dinger equation for charmonia and bottomonia \cite{Dumitru:2009ni}, where the masses of the charm and bottom quarks are taken to be  $m_c=1.3\, {\rm GeV}$ and $m_b=4.7\, {\rm GeV}$, respectively.

%%%%%%%%%%%%%%%%%%%%%%%%%%%%%%%%%%%%%%%%%%%%%%%%%%%%%%%%%%%%
\section{3D Anisotropic Potential Model}\label{sec-3}
%%%%%%%%%%%%%%%%%%%%%%%%%%%%%%%%%%%%%%%%%%%%%%%%%%%%%%%%%%%%%%
The real and imaginary part of the 3D anisotropic potential model as derived in our previous work~\cite{Dong:2022mbo} are
%%%
\begin{equation}
\mathrm{Re}\,V_{\mathrm{Aniso}} (r, \theta, \xi) = \alpha m_D^A \left(\frac{1-e^{-r m_D^R}}{r m_D^R}\right)-\alpha m_D^A - \frac{\sigma}{m_D^A}\left(2 + r m_D^R\right) e^{-r m_D^R}  + \frac{2\sigma}{m_D^A}-\frac{\alpha}{r}\, ,
\end{equation}
%%%
	
	%%%
	\begin{equation}
	\mathrm{Im}\,V_{\mathrm{Aniso}} (r, \theta, \xi) = \alpha \lambda^A \phi_2 \left(r m_D^I\right) - \alpha \lambda^A - \frac{8\sigma \lambda^A}{\left(m_D^A\right)^2}\phi_3 \left(r m_D^I\right) +  \frac{24\sigma \lambda^A}{\left(m_D^A\right)^2}\phi_4 \left(r m_D^I\right) -\frac{4\sigma \lambda^A}{\left(m_D^A\right)^2} \, ,
	\end{equation}
	%%%
	where,
	%%%
	\begin{equation}
	m_D^A =m_D \left(1-\frac{\xi}{6}\right)\, ,\quad \lambda^A = \lambda\left(1-\frac{\xi}{6}\right) ,\label{p15}
	\end{equation}
	%%%
	and
	%%%
	\begin{equation}
	m^R_D = m_D \bigg[1+\xi\left(0.108\cos 2\theta -0.131\right)\bigg]\, , \quad m^I_D =m_D \bigg[1+\xi\left(0.026\cos 2\theta -0.158\right)\bigg] .\label{p16}
	\end{equation}
	%%%
Eq(\ref{p15}) assures a correct asymptotic behavior of the potential. Eq. (\ref{p16}) was obtained by matching effective and exact result at ${\hat r}= 1$ as described in ~\cite{Dong:2022mbo}.
%%%%%%%%%%%%%%%%%%%%%%%%%%%%%%%%%%%%%%%%%%%%%%%%%%%%%%%%%%%%
\section{1D Effective Potential Model}\label{sec-4}
%%%%%%%%%%%%%%%%%%%%%%%%%%%%%%%%%%%%%%%%%%%%%%%%%%%%%%%%%%%%%%
Due to the angular dependence in the 3D anisotropic potential model, solving a 3D Schr{\"o}dinger equation to find various in-medium properties of the quarkonium states is rather time consuming and much more complicated. One possible solution to this problem is to introduce an angle-averaged effective screening mass ${\cal{M}}_{l m}(\lambda,\xi)$ ~\cite{Dong:2021gnb}
\ba\la{effm0}
{\cal{M}}^{R,I}_{l m}(\lambda,\xi)&=&\langle {\rm{Y}}_{l m}(\theta,\phi)| m^{R,I}_D(\lambda,\xi,\theta) | {\rm{Y}}_{l m}(\theta,\phi)\rangle\, ,\nonumber \\
&=&\int_{-1}^{1} d \cos \theta \int_{0}^{2\pi} d \phi {\rm{Y}}_{l m}(\theta,\phi)  m^{R,I}_D(\lambda,\xi,\theta) {\rm{Y}}^*_{l m}(\theta,\phi)\, ,
\ea
and where ${\rm{Y}}_{l m}(\theta,\phi)$ refers to the spherical harmonics with azimuthal quantum number $l$ and magnetic quantum number $m$. The main advantage of using an angle-averaged effective screening mass ${\cal{M}}_{l m}(\lambda,\xi)$ is to utilize the spherical symmetry in the potential model which significantly simplifies the numerics.

The real and imaginary part of the 1D effective potential model as derived in our previuos work~\cite{Dong:2022mbo} are
%%%
\begin{equation}
\mathrm{Re}\,V_{\mathrm{Eff}} (r,\xi) = \alpha m_D^A \left(\frac{1-e^{-r {\cal{M}}_{lm}^R}}{r {\cal{M}}_{lm}^R}\right)-\alpha m_D^A - \frac{\sigma}{m_D^A}\left(2 + r {\cal{M}}_{lm}^R\right) e^{-r {\cal{M}}_{lm}^R}  + \frac{2\sigma}{m_D^A}-\frac{\alpha}{r} \, ,
\end{equation}
%%%

%%%
\begin{equation}
\mathrm{Im}\,V_{\mathrm{Eff}} (r,\xi) = \alpha \lambda^A \phi_2 \left(r {\cal{M}}_{lm}^I\right) - \alpha \lambda^A - \frac{8\sigma \lambda^A}{\left(m_D^A\right)^2}\phi_3 \left(r {\cal{M}}_{lm}^I\right) +  \frac{24\sigma \lambda^A}{\left(m_D^A\right)^2}\phi_4 \left(r {\cal{M}}_{lm}^I\right) -\frac{4\sigma \lambda^A}{\left(m_D^A\right)^2} \, ,
\end{equation}
%%%
where,
%%%
\begin{equation}
K_{lm} = \frac{2l(l+1)-2 m^2 -1}{4l(l+1)-3} \,
\end{equation}
%%%
and
%%%
\begin{equation}
{\cal{M}}_{lm}^R = m_D \bigg[1+\xi\left(0.216 K_{lm} -  0.239\right)\bigg]~,\quad {\cal{M}}_{lm}^I = m_D \bigg[1+\xi\left(0.052 K_{lm} -  0.184\right)\bigg] .
\end{equation}
%%%
The $l$ and $m$ values of various quarkonium states are given in Table \ref{tab:01}.
%%%%%%%%%
\begin{table}[t]
	\centerline{\includegraphics[scale=.7]{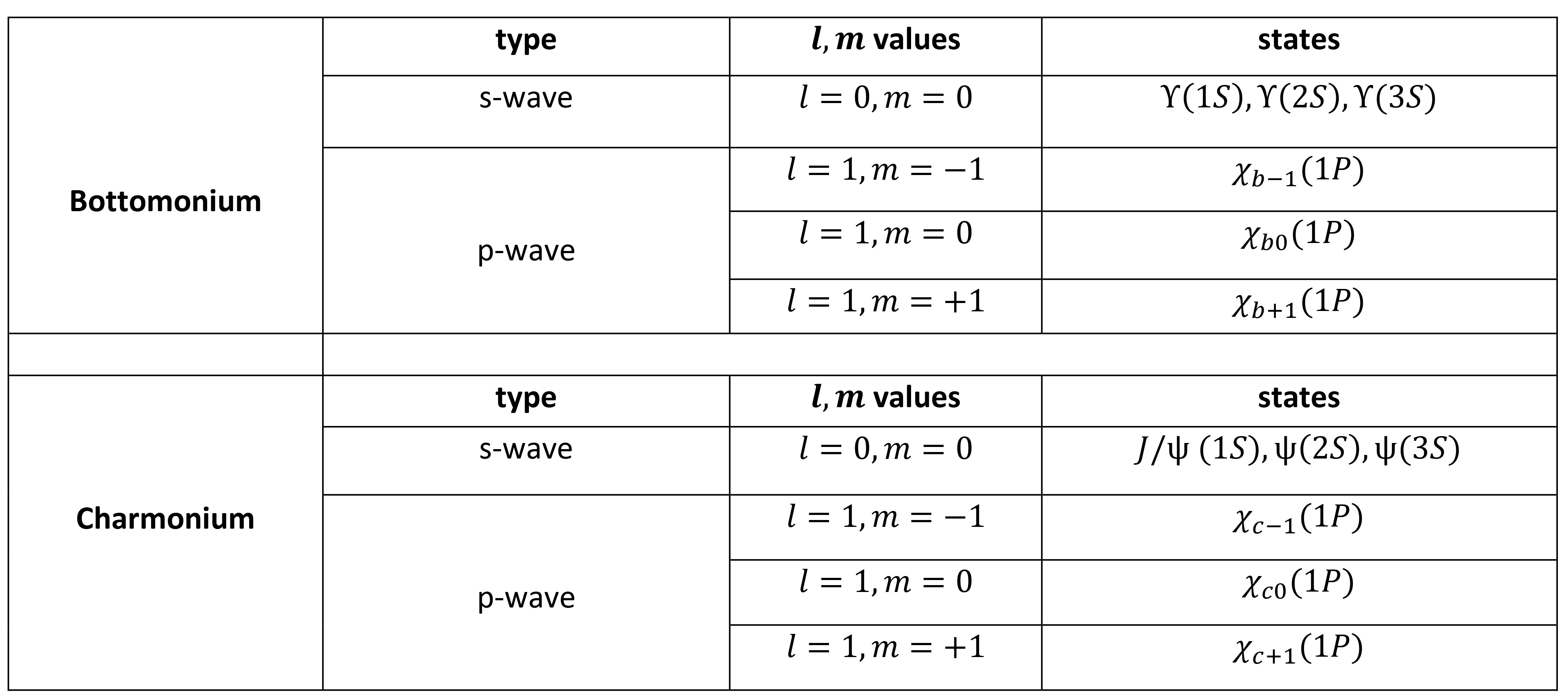}}
	\caption{$l$ and $m$ values of various quarkonium states.}
	\label{tab:01}
\end{table}
%%%%%%%%%%%%

%%%%%%%%%%%%%%%%%%%%%%%%%%%%%%%%%%%%%%%%%%%%%%%%%%%%%%%%%%%%
\section{Static Results}\label{sec-5}
%%%%%%%%%%%%%%%%%%%%%%%%%%%%%%%%%%%%%%%%%%%%%%%%%%%%%%%%%%%%%%
For the static solutions, we used a previously developed 3D eigensolver called quantumFDTD~\cite{Strickland:2009ft,Delgado:2020ozh}. Using this code, we compared results obtained with the 1D effective potential and the full 3D anisotropic potential. In Table~\ref{xi1}, we list the exact results of the eigenenergies (${\rm Re}\,E $), decay widths ($ {\rm Im}\,E$) and the binding energies ($E_{\rm bind}$) with the anisotropy parameter $\xi=1$ for $\Upsilon(1S)$ and $J/\Psi$.

In the numerical evaluations, we took $\alpha=0.272$ and  $\sigma=0.215\,{\rm GeV}^2$. For the $\Upsilon(1S)$ state, we used a lattice size of $N^3=512^3$ with a lattice spacing of $a=0.020\, {\rm GeV}^{-1}\approx 0.004 \,{\rm fm}$ giving a lattice size of $L=Na\approx 2.05\, {\rm fm}$. For the $J/\psi$, we used a lattice size of $N^3=256^3$ with a lattice spacing of $a=0.085\, {\rm GeV}^{-1}\approx 0.017\, {\rm fm}$ giving a lattice size of $L=Na\approx 4.35 \, {\rm fm}$.

%%%%%%%%%%%%%%%%%%%%%%%
\begin{table}[htpb]
\begin{center}
\setlength{\tabcolsep}{5mm}{
\begin{tabular}{  c  c  c  c c c }
\toprule[1pt]
$    \Upsilon(1S)     $  & ${\rm Re} E $  & $ \delta{\rm Re} E$ & $E_{\rm bind}$ &  $  {\rm Im} E  $ & $ \delta {\rm Im} E $ \\ \hline
$       T_o       $  & $182.869 $  &  $ 0.611  $ & $-662.669   $  &  $ 11.838     $ & $ 0.027 $   \\
$    1.1T_o       $  & $174.957 $  &  $ 0.593  $ & $-570.612   $  &  $ 14.830     $ & $ 0.031 $   \\
$    1.2T_o       $  & $166.556 $  &  $ 0.573  $ & $-493.689   $  &  $ 18.190     $ & $ 0.034 $   \\
$    1.4T_o       $  & $148.439 $  &  $ 0.531  $ & $-372.540   $  &  $ 26.004     $ & $ 0.039 $     \\
\bottomrule[1pt]	
\end{tabular}
\vspace{0.2cm}
\vspace{0.2cm}
\begin{tabular}{  c  c  c  c c c}
\toprule[1pt]
$      J/\Psi       $  & ${\rm Re} E$   & $ \delta{\rm Re} E$ &  $E_{\rm bind}$  & ${\rm Im} E$   & $ \delta{\rm Im} E $ \\ \hline
$       T_o         $  & $439.336 $   & $1.230 $ & $ -406.202 $   & $41.980$ & $ 0.107  $  \\
$    1.1T_o         $  & $422.207 $   & $1.163 $ & $ -323.362 $   & $51.467$ & $ 0.105  $  \\
$    1.2T_o         $  & $404.597 $   & $1.095 $ & $ -255.648 $   & $61.698$ & $ 0.098  $  \\
$    1.3T_o         $  & $386.604 $   & $1.028 $ & $ -199.583 $   & $72.564$ & $ 0.086  $  \\
$    1.4T_o         $  & $368.301 $   & $0.963 $ & $ -152.678 $   & $83.958$ & $ 0.070  $  \\
\bottomrule[1pt]	
\end{tabular}
}
\end{center}
\caption{The exact 3D results of the complex eigenenergies ($E$) and binding energies ($E_{\rm bind}$) for different quarkonium states at various temperatures with $\xi=1$. $\delta E$ are the differences in results obtained using 1D effective and 3D anisotropic potentials. Here $T_o$ is $192\,{\rm {MeV}}$ and all results are in MeV~\cite{Dong:2022mbo}.}
\label{xi1}
\end{table}
%%%%%

%%%%%%%%%%%%%%%%%%%%%%%%%%%%%%%%%%%%%%%%%%%%%%%%%%%%%%%%%%%%
\section{Dynamical Results}\label{sec-6}
%%%%%%%%%%%%%%%%%%%%%%%%%%%%%%%%%%%%%%%%%%%%%%%%%%%%%%%%%%%%%%

In order to solve the 3D Schr\"odinger equation in real time, we used a split-step pseudospectral method~\cite{Taha:1984jz} with temporal step size $\Delta t = 0.001$ fm/c. Once again we compare results obtained with the full 3D anisotropic potential to those obtained with the 1D effective potential. We evolve the wave function from $\tau = 0$ fm/c to $\tau = 0.25$ fm/c in the vacuum ($T=0$).  Starting at $\tau = \tau_0 = 0.25$ fm/c, we consider a fixed anisotropy parameter $\xi = 1$ and boost-invariant Bjorken evolution for the hard scale
%%%
\be
\lambda(\tau) = \lambda_0 \left( \frac{\tau_0}{\tau} \right)^{1/3} \, .
\ee
%%%
Here we take the initial hard scale to be $\lambda_0 =$ 630 MeV. Further details of the numerical method can be found in ~\cite{Dong:2022mbo}.

%%%%%%%%%%%%%%%%%%%%%%%%%%
\subsection{Bottomonium}
%%%%%%%%%%%%%%%%%%%%%%%%%%

\begin{figure}[t]
\centering
\includegraphics[width=1\textwidth]{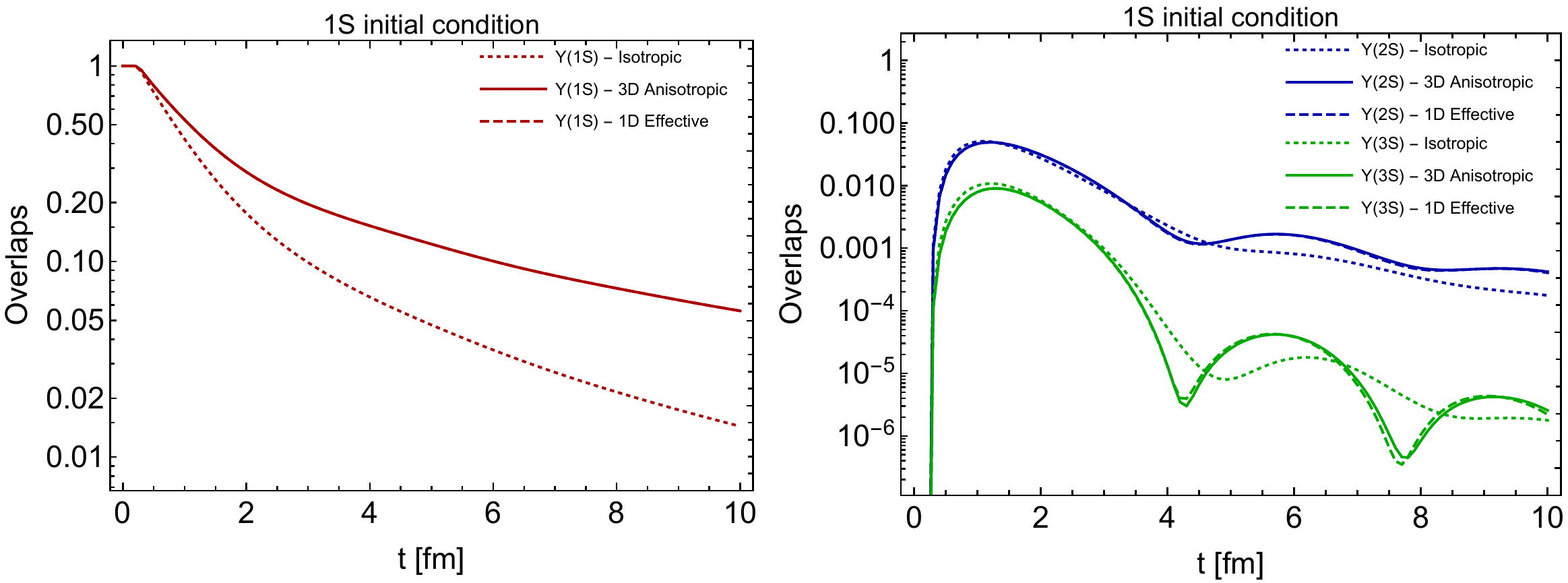}\\[1em]
\includegraphics[width=1\textwidth]{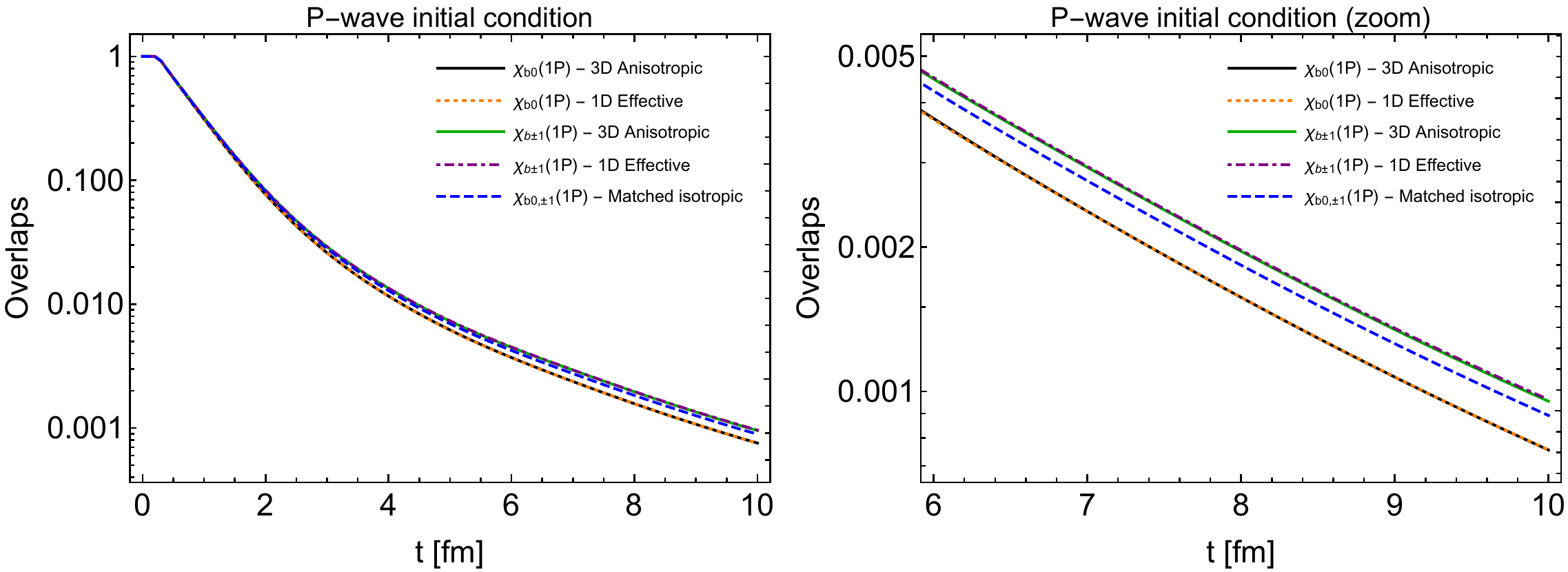}
\caption{The top row shows the overlaps of $\Upsilon(1S)$, $\Upsilon(2S)$, and $\Upsilon(3S)$ resulting from real-time solution of the Schr\"odinger equation. Here we initialized the wave function as pure $\Upsilon(1S)$ eigenstate. The bottom row shows the time evolution of the bottomonium p-wave overlaps resulting from initialization with different p-wave polarizations~\cite{Dong:2022mbo}.}
\label{fig:overlaps-swave}
\end{figure}

For bottomonium states we take the box size to be $L =$ 2.56 fm, $m_b = 4.7$ GeV, and use $N=128$ lattice points in each direction. The top row of fig.~\ref{fig:overlaps-swave} shows the time evolution of overlaps of the $\Upsilon(1S)$, $\Upsilon(2S)$, and $\Upsilon(3S)$ using a pure $\Upsilon(1S)$ eigenstate as the initial condition. Whereas the bottom row shows the time evolution of the bottomonium p-wave overlaps resulting from initialization with different p-wave polarizations. Results with pure $\Upsilon(2S)$ and $\Upsilon(3S)$ eigenstates and a Gaussian as the initial condition can be found in Ref.~\cite{Dong:2022mbo}.
%%%
%%%

%%%%%%%%%%%%%%%%%%%%%%%%%%
\subsection{Charmonium}
\begin{figure}
\centering
\includegraphics[width=1\textwidth]{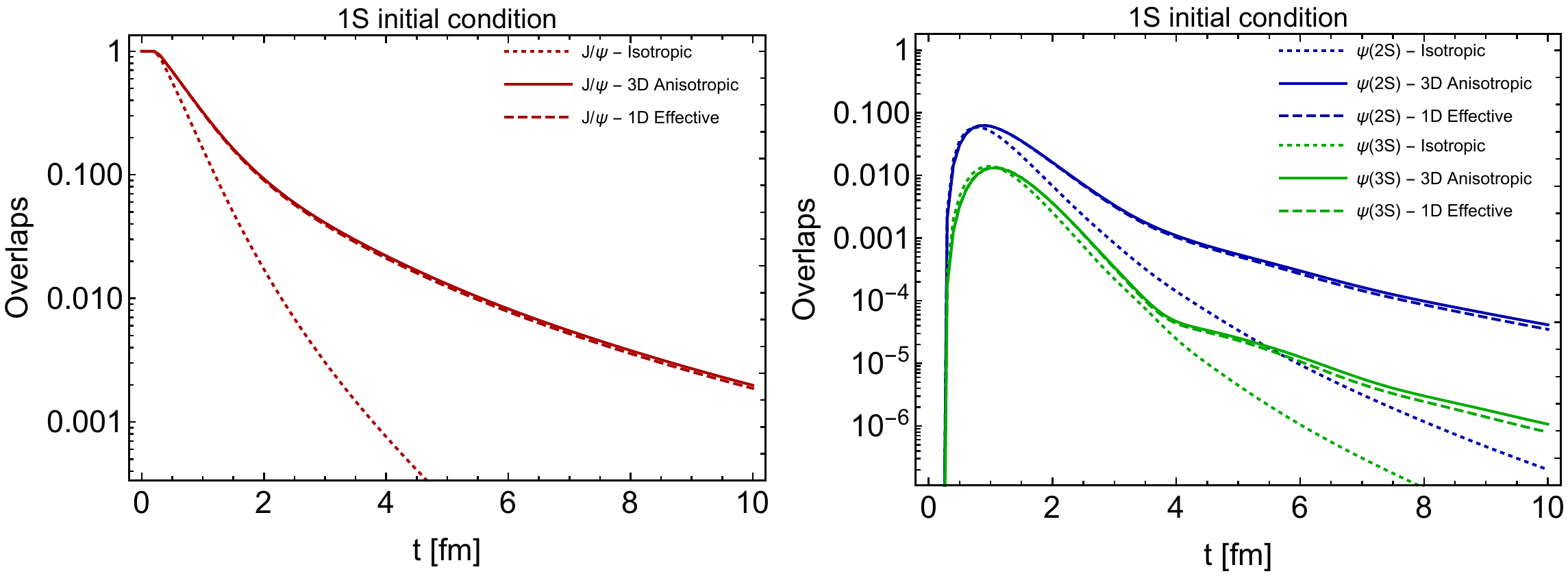}\\[1em]
\includegraphics[width=1\textwidth]{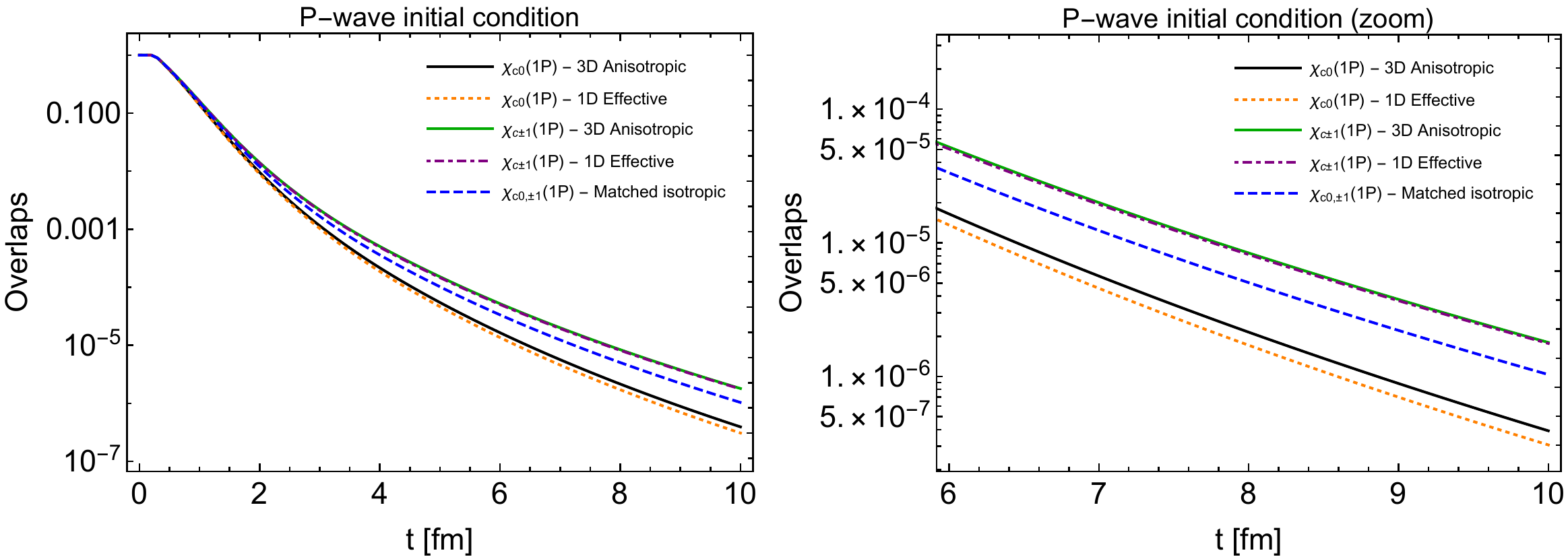}
\caption{The top row shows the overlaps of $J/\psi$, $\psi(2S)$, and $\psi(3S)$ resulting from real-time solution of the Schr\"odinger equation. Here we initialized the wave function as pure $J/\psi$ eigenstate. The bottom row shows the time evolution of the charmonium p-wave overlaps resulting from initialization with different p-wave polarizations~\cite{Dong:2022mbo}.}
\label{fig:overlaps-jpsi}
\end{figure}

%%%%%%%%%%%%%%%%%%%%%%%%%%
For charmonium states we take $L = $ 5.12 fm, $m_c = 1.3$ GeV, and use $N=128$ lattice points in each direction. The top row of the fig.~\ref{fig:overlaps-jpsi} shows the time evolution of overlaps of the $J/\psi$, $\psi(2S)$, and $\psi(3S)$ by using pure $J/\psi$ eigenstate as the initial condition. Whereas the bottom row shows the time evolution of the charmonium p-wave overlaps resulting from initialization with different p-wave polarizations. The results with pure $\psi(2S)$ and $\psi(3S)$ eigenstate and Gaussian as the initial condition can be found in Ref.~\cite{Dong:2022mbo}.
%%%
%%%

%%%%%%%%%%%%%%%%%%%%%%%%%%%%%%%%%%%%%%%%%%%%%%%%%%%%%%%%%%%%
\section{Conclusions}
%%%%%%%%%%%%%%%%%%%%%%%%%%%%%%%%%%%%%%%%%%%%%%%%%%%%%%%%%%%%%%

We have reduced anisotropic heavy-quark potentials to isotropic ones by introducing an effective screening mass that depends on the quantum numbers $l$ and $m$ of a given state. We demonstrated that, using the resulting 1D effective potential model, one can reproduce the full 3D results for the energies and binding energies of low-lying heavy-quarkonium bound states to relatively high accuracy.  This finding is important because it can be used to incorporate anisotropy effects into one-dimensional real-time Schr\"odinger equations which underpin phenomenological calculations of bottomonium suppression in open quantum systems approaches.

%%%%%%%%%%%%%%%%%%%%%%%%%%%%%%%%%%%%%%%%%%%%%%%%%%%%%%%%%%%%
\section*{Acknowledgements}
%%%%%%%%%%%%%%%%%%%%%%%%%%%%%%%%%%%%%%%%%%%%%%%%%%%%%%%%%%%%%%

The speaker (A.I.) would like to thank the organizers of the Quark Confinement and the Hadron Spectrum conference 2022 in Stavanger, Norway for the opportunity to present this talk. The work of Y.G. is supported by the NSFC of China under Project No. 12065004 and 12147211. M.S. and A.I.  were supported by the U.S. Department of Energy, Office of Science, Office of Nuclear Physics Award No.~DE-SC0013470. A.R. gladly acknowledges support from the Research Council of Norway under the FRIPRO Young Research Talent grant 286883 and from UNINETT Sigma2 - the National Infrastructure for High Performance Computing and Data Storage in Norway under project NN9578K-QCDrtX "Real-time dynamics of nuclear matter under extreme conditions".

%%%%%%%%%%%%%%%%%%%%%%%%%%%%%%%%%%%%%%%%%%%%%%%%%%%%%%%%%%%%%%%%%%%%%%
\bibliography{references}

\begin{thebibliography}{29}

\bibitem{Matsui:1986dk}
T.~Matsui, H.~Satz, Phys. Lett. B \textbf{178}, 416 (1986)

\bibitem{Karsch:1987pv}
F.~Karsch, M.T. Mehr, H.~Satz, Z. Phys. C \textbf{37}, 617 (1988)

\bibitem{Laine:2006ns}
M.~Laine, O.~Philipsen, P.~Romatschke, M.~Tassler, JHEP \textbf{03}, 054
  (2007), \texttt{hep-ph/0611300}

\bibitem{Brambilla:2008cx}
N.~Brambilla, J.~Ghiglieri, A.~Vairo, P.~Petreczky, Phys. Rev. D \textbf{78},
  014017 (2008), \texttt{0804.0993}

\bibitem{Beraudo:2007ky}
A.~Beraudo, J.P. Blaizot, C.~Ratti, Nucl. Phys. A \textbf{806}, 312 (2008),
  \texttt{0712.4394}

\bibitem{Escobedo:2008sy}
M.A. Escobedo, J.~Soto, Phys. Rev. A \textbf{78}, 032520 (2008),
  \texttt{0804.0691}

\bibitem{Brambilla:2010vq}
N.~Brambilla, M.A. Escobedo, J.~Ghiglieri, J.~Soto, A.~Vairo, JHEP \textbf{09},
  038 (2010), \texttt{1007.4156}

\bibitem{Brambilla:2011sg}
N.~Brambilla, M.A. Escobedo, J.~Ghiglieri, A.~Vairo, JHEP \textbf{12}, 116
  (2011), \texttt{1109.5826}

\bibitem{Brambilla:2013dpa}
N.~Brambilla, M.A. Escobedo, J.~Ghiglieri, A.~Vairo, JHEP \textbf{05}, 130
  (2013), \texttt{1303.6097}

\bibitem{Dumitru:2009ni}
A.~Dumitru, Y.~Guo, A.~Mocsy, M.~Strickland, Phys. Rev. D \textbf{79}, 054019
  (2009), \texttt{0901.1998}

\bibitem{Margotta:2011ta}
M.~Margotta, K.~McCarty, C.~McGahan, M.~Strickland, D.~Yager-Elorriaga, Phys.
  Rev. D \textbf{83}, 105019 (2011), [Erratum: Phys.Rev.D 84, 069902 (2011)],
  \texttt{1101.4651}

\bibitem{Strickland:2011mw}
M.~Strickland, Phys. Rev. Lett. \textbf{107}, 132301 (2011), \texttt{1106.2571}

\bibitem{Strickland:2011aa}
M.~Strickland, D.~Bazow, Nucl. Phys. A \textbf{879}, 25 (2012),
  \texttt{1112.2761}

\bibitem{Krouppa:2015yoa}
B.~Krouppa, R.~Ryblewski, M.~Strickland, Phys. Rev. C \textbf{92}, 061901
  (2015), \texttt{1507.03951}

\bibitem{Krouppa:2016jcl}
B.~Krouppa, M.~Strickland, Universe \textbf{2}, 16 (2016), \texttt{1605.03561}

\bibitem{Krouppa:2017jlg}
B.~Krouppa, A.~Rothkopf, M.~Strickland, Phys. Rev. D \textbf{97}, 016017
  (2018), \texttt{1710.02319}

\bibitem{Islam:2020gdv}
A.~Islam, M.~Strickland, Phys. Lett. B \textbf{811}, 135949 (2020),
  \texttt{2007.10211}

\bibitem{Islam:2020bnp}
A.~Islam, M.~Strickland, JHEP \textbf{21}, 235 (2020), \texttt{2010.05457}

\bibitem{Wen:2022yjx}
L.~Wen, B.~Chen (2022), \texttt{2208.10050}

\bibitem{Romatschke:2003ms}
P.~Romatschke, M.~Strickland, Phys. Rev. D \textbf{68}, 036004 (2003),
  \texttt{hep-ph/0304092}

\bibitem{Strickland:2014pga}
M.~Strickland, Acta Phys. Polon. B \textbf{45}, 2355 (2014), \texttt{1410.5786}

\bibitem{Berges:2020fwq}
J.~Berges, M.P. Heller, A.~Mazeliauskas, R.~Venugopalan, Rev. Mod. Phys.
  \textbf{93}, 035003 (2021), \texttt{2005.12299}

\bibitem{Thakur:2013nia}
L.~Thakur, U.~Kakade, B.K. Patra, Phys. Rev. D \textbf{89}, 094020 (2014),
  \texttt{1401.0172}

\bibitem{Dong:2021gnb}
L.~Dong, Y.~Guo, A.~Islam, M.~Strickland, Phys. Rev. D \textbf{104}, 096017
  (2021), \texttt{2109.01284}

\bibitem{Dong:2022mbo}
L.~Dong, Y.~Guo, A.~Islam, A.~Rothkopf, M.~Strickland, JHEP \textbf{09}, 200
  (2022), \texttt{2205.10349}

\bibitem{Guo:2018vwy}
Y.~Guo, L.~Dong, J.~Pan, M.R. Moldes, Phys. Rev. D \textbf{100}, 036011 (2019),
  \texttt{1806.04376}

\bibitem{Strickland:2009ft}
M.~Strickland, D.~Yager-Elorriaga, J. Comput. Phys. \textbf{229}, 6015 (2010),
  \texttt{0904.0939}

\bibitem{Delgado:2020ozh}
R.L. Delgado, S.~Steinbei\ss{}er, M.~Strickland, J.H. Weber, Comput. Phys.
  Commun. \textbf{272}, 108250 (2022), \texttt{2006.16935}

\bibitem{Taha:1984jz}
T.R. Taha, M.J. Ablowitz, J. Comput. Phys. \textbf{55}, 203 (1984)

\end{thebibliography}
%%%%%%%%%%%%%%%%%%%%%%%%%%%%%%%%%%%%%%%%%%%%%%%%%%%%%%%%%%%%%%%%%%%%%%%%

\end{document}